\documentclass[12pt,a4paper]{article}
\usepackage{type1cm,amsmath,graphicx,indentfirst}
\usepackage[psamsfonts]{amssymb}
\numberwithin{equation}{section}

\newcommand {\g}{\mathfrak g}
\newcommand {\sub}{\mathfrak a}
\newcommand {\del}{\partial}
\newcommand {\Ad}{\text{Ad}}

\newcommand{\tr}{\text{tr}}

\begin{document}
%%% Title page %%%%%
\begin{titlepage}

 \renewcommand{\thefootnote}{\fnsymbol{footnote}}
\begin{flushright}
 \begin{tabular}{l}
 %DESY 10-098\\
 %WITS-CTP-54%\\
% arXiv:1103.xxxx\\ %This should be replaced after submittion.
% \today %This should be commented out.
 \end{tabular}
\end{flushright}

 \vfill
 \begin{center}

% \vskip 2.5 truecm

\noindent{\large \textbf{Supergroup - extended super Liouville correspondence
}}\\
\vspace{1.5cm}

\noindent{ Thomas Creutzig,$^{a,\,b}$\footnote{E-mail: tcreutzig@mathematik.tu-darmstadt.de} Yasuaki Hikida$^c$\footnote{E-mail:
hikida@phys-h.keio.ac.jp} and Peter B. R\o nne$^{d,\, e}$\footnote{E-mail: peter.roenne@uni-koeln.de}}
\bigskip

 \vskip .6 truecm
\centerline{\it $^a$Department of Physics and Astronomy, University of North Carolina,}
\centerline{\it Phillips Hall, CB 3255, Chapel Hill, NC 27599-3255, USA}
\medskip
\centerline{\it $^b$Fachbereich Mathematik,
Technische Universit\"{a}t Darmstadt,}
\centerline{\it Schlo\ss gartenstr. 7
64289 Darmstadt, Germany}
\medskip
\centerline{\it $^c$Department of Physics, and Research and Education
Center for Natural Sciences,}
\centerline{\it  Keio University, Hiyoshi, Yokohama 223-8521, Japan}
\medskip
\centerline{\it $^d$Institut f\"{u}r Theoretische Physik, Universit\"{a}t zu K\"{o}ln,
} \centerline{\it
Z\"{u}lpicher Stra{\ss}e 77, 50937 Cologne, Germany}
\medskip
\centerline{\it $^e$National Institute for Theoretical Physics and Centre
for Theoretical Physics,} \centerline{\it University of the Witwatersrand, Wits, 2050, South Africa}

 \vskip .4 truecm

 \end{center}

 \vfill
\vskip 0.5 truecm

\begin{abstract}

We derive a relation between correlation functions of supergroup WZNW models
and conformal field theories with extended superconformal symmetry.
The supergroups considered have a bosonic subgroup of the form $SL(2) \times A$
for some Lie group $A$. The corresponding conformal field theory is a super Liouville
field theory coupled with the WZNW model on $A$.
An example is a correspondence between the PSU(1,1$|$2) WZNW model and small $\mathcal N=4$ super Liouville field theory.
The OSP($n|$2) WZNW model is related to a superconformal field theory with SO$(n)$ extended superconformal symmetry of the Knizhnik-Bershadsky type. In the case $n=4$ this is
simply the large $\mathcal N=(4,4)$ superconformal symmetry.
Besides these two examples we make a general derivation encompassing the WZNW models on supergroups SL(2$|n$), D(2,1;$\alpha$),
OSP(4$|2n$), F(4) and G(3) and their relation to models with
extended superconformal algebras as symmetry.
%The exceptional case  D(2,1;$\alpha$) is a continuous family interpolating between large and small $\mathcal N=4$ superconformal symmetry.

\end{abstract}
\vfill
\vskip 0.5 truecm

\setcounter{footnote}{0}
\renewcommand{\thefootnote}{\arabic{footnote}}
\end{titlepage}

\newpage

\tableofcontents
%%%%%%%%%%%%%%%%%%%%%%%%%%%%%%%%%%%%%%%%%%%%%%%%%%%%%%%%%%%%%%%%%%%%%%
%\newpage

\section{Introduction}

Recently, supergroup models have been studied by many researchers
due to the development of the AdS/CFT correspondence \cite{Maldacena}.
The correspondence relates superstring theories on Anti-de Sitter
(AdS) spaces with conformal field theories on the boundary, and
it is known that the superstring theories are described by
models on supergroups/cosets. For example, superstring theories on
AdS$_5 \times$S$^5$ and AdS$_3 \times$S$^3$ are described
by utilizing  PSU$(2,2|4)$ \cite{MT}
and PSU$(1,1|2)$ symmetry \cite{BVW}, respectively.
Supergroup models are also used to
describe disordered systems in the context of condensed matter
physics \cite{Efetov}. Especially interesting in this respect are the models proposed for the plateau transition in the integer quantum Hall effect \cite{Zirnbauer:1999ua} (see also \cite{Bhaseen:1999nm,Guruswamy:1999hi}).

In this note, we study a simple type of supergroup models, namely,
Wess-Zumino-Novikov-Witten (WZNW) models based on supergroups.
These have been studied in the bulk \cite{Rozansky:1992rx, Schomerus:2005bf,GQS,Saleur:2006tf,Quella:2007hr} and with boundaries \cite{Creutzig:2007jy,Creutzig:2008ek,Creutzig:2008ag,Creutzig:2008an,Creutzig:2009zz,Creutzig:2010zp}.
One way to deal with a supergroup WZNW model is to rewrite it as
a WZNW model on the bosonic subgroup coupled with free fermions
\cite{Quella:2007hr}. However, this method can be applied
only to type I supergroups.
Another way is to utilize the relation to super Liouville theory
developed in \cite{HS2}, where the structure constants of
OSP$(1|2)$ WZNW model were computed.
In the paper, the relation between correlation functions of
OSP$(n|2)$ WZNW model and ${\cal N}=n$ super Liouville theory
was obtained for $n=1,2$, which is a generalization of the
relation between correlation functions of SL(2) WZNW model
and Liouville field theory \cite{RT,HS}.
The aim of this note is to extend the relation to more generic cases,
where the Liouville theory side has extended superconformal symmetry.

The extension of the conformal symmetry to ${\cal N} = n \leq 4$
supersymmetry was found
in \cite{Ademollo},
and it has SU(2) or SO($n$) symmetry.
A theory with this type of extended superconformal symmetry can
be realized by a Liouville-like theory coupled with SU(2) or SO($n$) WZNW models
\cite{Ivanov:1987mz,IKL}.
In this case, the (anti-)commutators of generators of
the superconformal symmetry are given by linear combinations of
the generators themselves.
If we do not require such a linear relation,
the superconformal symmetry can be extended
with U$(n)$ or SO$(n)$ symmetry for arbitrary $n$ as shown by Knizhnik and Bershadsky \cite{K,B}.
In this paper we call this type of symmetry ``superconformal $W$-algebra''
since the (anti-)commutators may lead to products of generators.
In fact, the relation between supergroup WZNW models and super
Liouville field theory is not entirely new.
In \cite{BO} it was shown that utilizing quantum Hamiltonian
reduction OSP$(n|2)$ WZNW models reduce to ${\cal N}=n$ super
Liouville field theory.
This analysis was extended
in \cite{IM, Ito:1992ig, Ito:1992hv,Ito:1992bi,Ito:1993qi} to
supergroups whose bosonic subgroup is of the form of
SL(2)$\times A$. The extended superconformal algebras were also
classified by an algebraic approach in \cite{Bowcock:1992bm,Fradkin:1992bz}.

A strong motivation to study these relations comes from the proof of the Fateev-Zamolodchikov-Zamolodchikov (FZZ) duality \cite{FZZ,Hikida:2008pe}. The FZZ duality relates  Witten's semi-infinite cigar model, given by the coset $H_3^+/\mathbb{R}$, to sine-Liouville field theory. In the proof the main ingredients were the above mentioned relation of the SL(2) WZNW model to Liouville field theory together with the self duality of Liouville theory. The proof was generalized in \cite{Creutzig:2010bt} to the case with branes in the cigar and to the supersymmetric case where the relation takes the form of a mirror duality
\cite{Hori:2001ax}. 
In the proofs it was essential that the relation was known in detail directly relating correlators to correlators. It would be interesting to find similar dualities for the supergroups that we consider in this paper. 
One example may be given by $\text{OSP}(1|2)/\text{U}(1)$ model discussed in 
\cite{Giribet:2009eb}.
The first step is taken by making the relation to the superconformal Liouville field theories precise using the path integral formalism.

Our strategy is as follows. We consider supergroups whose bosonic subgroups are of the form SL(2)$\times A$.
The idea is to use the path integral techniques from \cite{HS} to integrate out the fields corresponding to the roots of SL(2). This requires finding a suitable free field realization of the WZNW model such that these fields do not appear in
the interaction terms. We find that this can be achieved using a special five-decomposition
of the Lie superalgebra. Then we can proceed as in \cite{HS,HS2}, 
that is after integrating out the fields corresponding to the roots of SL(2) and performing certain field redefinitions we get a theory endowed with extended superconformal symmetry. Due to the field redefinitions involved, an $N$-point correlation function in the WZNW model on a sphere
is equal to a $(2N-2)$-point correlator with $N-2$ degenerate fields inserted.

The organization of this note is as follows:
In the next section we study the relation to Liouville field
theory with ${\cal N} \leq 4$ superconformal symmetry. First we relate the PSU$(1,1|2)$ WZNW model
to Liouville theory with small ${\cal N}=(4,4)$ superconformal symmetry in subsection \ref{small},
and we consider OSP$(n|2)$ WZNW models
which relate to Liouville theory with ${\cal N}=(n,n)$
superconformal symmetry  in subsection \ref{large}.
In section \ref{general} we consider the general case
when the supergroup has its bosonic sub-algebra of the form
SL(2)$\times A$.
Notations for the Lie superalgebras are
given in appendix \ref{notation}, and the symmetries of the reduced theories
are studied in appendix \ref{symmetry}.

\section{WZNW models and super Liouville theories}

In this section we study the relation between supergroup WZNW models and Liouville
field theory with superconformal and extended superconformal symmetry.
We start by deriving a correspondence between the PSU$(1,1|2)$ WZNW model
and Liouville theory with small ${\cal N}=(4,4)$ superconformal symmetry.
This case is relevant for the study of the
AdS$_3$/CFT$_2$ correspondence where string theory on AdS$_3\times$S$^3$ is related to the supergroup \cite{BVW,Creutzig:2010ne}. Further note that small ${\cal N}=(4,4)$ superconformal symmetry on the dual CFT side played an important role in the protection of certain correlators \cite{deBoer:2008ss}.
In subsection \ref{large} we then consider OSP$(n|2)$ WZNW models,
which are related to Liouville theory with ${\cal N}=(n,n)$
superconformal symmetry. It reproduces the result in
\cite{HS2} for $n=1,2$.
Moreover for $n>4$ the commutators of symmetry generators contain the products of 
generators and the symmetry algebra can thus be called as
superconformal $W$-algebra.

\subsection{From PSU(1,1$|$2) to small ${\cal N}=4$ Liouville theory}
\label{small}

We derive a correspondence between the correlators of
the WZNW model of the supergroup PSU(1,1$|$2) and
small ${\cal N}=4$ Liouville field theory.
The bosonic subgroup of the supergroup is the product of  $ \text{SU}(1,1)\simeq\text{SL}(2)$ and $\text{SU}(2)$.
To be more precise, we will here and in the following actually consider sigma models whose bosonic subspace instead of SL(2) contain the coset $H_3^+=\text{SL}(2,\mathbb{C})/\text{SU}(2)$  i.e. the space of hermitian elements of SL$(2,\mathbb{C})$. The Euclidean hyperbolic space $H_3^+$ can then be related to Liouville theory using the method of \cite{HS}. The action for the $H_3^+$ model can, however, be written as the WZNW action based on Hermitian matrices, see \cite{Gawedzki:1991yu}.

The derivation proceeds as follows: First we introduce
the WZNW model and express the correlation function as a path integral.
The fields corresponding to the roots of SL(2) are then integrated out. Finally we have to
rotate some fields to arrive at the desired small
${\cal N}=4$ Liouville field theory.

\subsubsection{The PSU(1,1$|$2) WZNW model}

The first step is to find an appropriate free field realization of the WZNW model.
The explicit form of the WZNW model action depends on the parametrization of the supergroup valued field.
For our purposes it is of crucial importance to choose this to  be $g = \alpha G \beta$ with
\begin{equation}
 \begin{split}\label{eq:parampsl22}
 \alpha\  &=\ \exp (\tfrac{1}{\sqrt2} (\theta^1_2 S^-_{12} + \theta^1_1 S^-_{22} ) )
 \exp (\tfrac{1}{\sqrt2} (\theta^2_1 S^-_{11} - \theta^2_2 S^-_{21} ))\exp(\gamma E_1^-) ~, \\
 \beta\ &=\  \exp (\tfrac{1}{\sqrt2} (\bar \theta^2_1 S^+_{12} - \bar \theta^2_2 S^+_{22} ))
 \exp (\tfrac{1}{\sqrt2} (\bar \theta^1_2 S^+_{11} + \bar \theta^1_1 S^+_{21} ))\exp(\bar \gamma E_1^+) ~, \\
G \ &=\   \exp( - 2 \phi E_1^0 ) \begin{pmatrix}
  \mathbb{I}_2 & 0 \\
  0 & q
\end{pmatrix} ~,
 \end{split}
\end{equation}
and $q$  a $2\times 2$ SU(2)-matrix.
The notation for the generators is summarized in appendix \ref{notation}.
This parametrization is different from the one chosen in \cite{GQS},
which is suitable for free fermion resolution of type I cases
\cite{Quella:2007hr}. Here we employ a decomposition of  PSU$(1,1|2)$ that resembles  type II
supergroups in order to apply the method in \cite{HS2}. A
more detailed explanation of the decompositions of the supergroups can be found in the next section.

Using the well-known Polyakov-Wiegmann identity, the action is
\begin{align}
 S^{\textrm{WZNW}}[g]_k = S^{\textrm{WZNW}}[q]_k + \frac{k}{2 \pi} \int d^2 z \Bigl[ \bar \partial \phi \partial \phi
  &+ e^{ - 2 \phi} (\bar \partial \gamma - \theta_2 \bar \partial \theta_1^t
) (\partial \bar \gamma + \bar \theta_2 \partial \bar \theta_1^t ) \nonumber \\
& + e^{ - \phi}  \partial \bar \theta_1 q^{-1} \bar  \partial \theta_1^t +
  e^{ - \phi}   \bar \partial \theta_2 q \partial \bar \theta_2^t  \Bigr]~,
   \label{psu22action}
\end{align}
where we have used
\begin{align}
\theta_1 =  (
   \theta^1_1 ,
   \theta^1_2 ) ~, \qquad
\bar \theta_1 = (\bar \theta^1_1 , \bar \theta^1_2)
~, \qquad
\theta_2 = (\theta^2_1 , \theta^2_2) ~, \qquad
\bar \theta_2 = (
  \bar \theta^2_1 ,
  \bar \theta^2_2 )
 ~.
\end{align}
The transpose operation was expressed as $(q^t)_{i,j} = q_{j,i}$.
The action can be rewritten in a first order form.
Introducing $\beta,\bar \beta$ and $p^a_b , \bar p^a_b$
with $a,b = 1,2$, we now have
\begin{align}
   S[g,\beta,\bar \beta,p , \bar p]_k\ &= \ S^{\textrm{WZNW}}[q]_{k-2} + S_0 + S_{\text{int}} ~, \\
S_0 \ &= \ \frac{1}{2 \pi} \int d^2 z \Bigl[
\bar \partial \phi \partial \phi + \frac{\hat Q}{4} \sqrt{g} {\cal R} \phi
- \beta \bar \partial \gamma - \bar \beta \partial \bar \gamma + \sum_{a = 1,2}
 ( p_a \bar \partial \theta_a^t + \bar p_a \partial \bar \theta_a^t )\Bigr]  ~, \nonumber \\
S_{\text{int}} \ &= \   \frac{1}{2 \pi} \int d^2 z \Bigl[ -\frac{1}{k} \beta \bar \beta e^{2 b \phi}  + \frac{1}{k} (p_1 - \beta \theta_2 ) q
 ( \bar p_1 + \bar \beta \bar \theta_2 )^t e^{ b \phi }
  + \frac{1}{k} \bar p_2 q^{-1} p_2^t e^{ b  \phi } \Bigr]  ~, \nonumber 
\end{align}
with $b = 1/\sqrt{k}$. Here the level of SU(2) has been shifted due to the change in the path integral measure (see also \cite{GQS}) and a background charge have been added $\hat Q = - b$.
Moreover, we have defined
\begin{align}
p_1 = (p^1_1 , p^1_2) ~, \qquad
\bar p_1 = (
  \bar p^1_1 ,
  \bar p^1_2 )
~, \qquad
p_2 = (
  p^2_1 ,
   p^2_2 ) ~, \qquad
\bar p_2 =(\bar p^2_1 ,\bar  p^2_2)
~.
\end{align}

In order to write down the vertex operators, it is convenient to bosonize the
fermionic fields as
\begin{align}
  p^a_b = e^{i Y^a_b} ~, \qquad \theta^a_b = e^{- i Y^a_b} ~, \qquad
  \bar p^a_b = e^{i \bar Y^a_b} ~, \qquad
  \bar \theta^a_b = e^{- i \bar Y^a_b} ~.
\end{align}
Then the vertex operators are
\begin{align}
 V^{s^a_b}_{j,L} (\mu | z) =  \mu ^{j + 1 + \frac{1}{2} s^a_b} \bar \mu ^{j + 1 + \frac{1}{2} \bar s^a_b}
 e^{ i s^a_b Y^a_b + i \bar s^a_b \bar Y^a_b}e^{\mu \gamma - \bar \mu \bar \gamma}
  e^{2b(j+1) \phi} V^\text{SU(2)}_L (q) ~,
\end{align}
where $L$ labels the representation of su(2).

\subsubsection{The correspondence with small $\mathcal N=4$ Liouville theory}

We now consider correlation functions. We first perform the path integral for the fields $\gamma,\bar\gamma$ and then
for $\beta,\bar\beta$. Then after a field redefinition the correlation function takes the form of those in small $\mathcal N=4$
Liouville theory with extra insertions of degenerate fields.

We consider correlation functions of the form
\begin{align}
 \left\langle \prod_{i=1}^N V^{ {s^a_b}_i }_{j_i,L_i} (\mu_i | z_i)  \right\rangle ~.
\end{align}
Following the analysis in \cite{HS} we integrate out first
$\gamma, \bar \gamma$ and then $\beta , \bar \beta$. After this
procedure, the field $\beta$ is replaced by the
function
\begin{align}
 \sum_{i=1}^N \frac{\mu_i}{z-z_i} = u \frac{\prod_{l=1}^{N-2} (z - y_l)}
 {\prod_{i=1}^N (z - z_i) } = u {\cal B} (y_l , z_i ; z) ~,
\end{align}
and similarly $\bar \beta$ by $ - u \bar {\cal B} (\bar y_l , \bar z_i ; \bar z)$.
We remove the function
${\cal B}$ from the action by shifting the fields $\phi,Y_b^a$ as
\begin{align}\label{eq:ShiftabsorbB}
 \phi + \frac{1}{2b} \ln |{\cal B}|^2 \to \phi ~, \qquad
 Y_b^a - \frac{i}{2} \ln {\cal B} \to Y_b^a ~.
\end{align}
Note, that this shift changes conformal dimension of the fermions to $1/2$.
Then we have (for a detailed derivation, see \cite{HS2})
\begin{align}
 \left\langle \prod_{i=1}^N V^{ {s^a_b}_i }_{j_i,L_i} (\mu_i | z_i)  \right\rangle_{\textrm{WZNW}}  = \delta^{(2)} (\sum_{i=1}^N \mu_i )
 | \Theta_N |^2  \left\langle \prod_{i=1}^N
V^{ {s^a_b}_i +1/2 }_{b(j_i + 1) + 1/2b,L_i}  ( z_i)
 \prod_{l=1}^{N-2} V^{ -1/2 }_{-1/2b,0} ( y_l)
 \right\rangle  ~.
\end{align}
Here $L=0$ denotes the identity representation of su$(2)$.
The right hand side is evaluated with the action
\begin{equation}\label{eq:N4lft}
\begin{split}
 S\ &=\  S^{\textrm{WZNW}}[q]_{k-2} + S_0 + S_{\text{int}} ~, \\
S_0 \ &= \ \frac{1}{2 \pi} \int d^2 z \Bigl[
\bar \partial \phi \partial \phi + \frac{Q}{4} \sqrt{g} {\cal R} \phi
+ \sum_{a  = 1,2}
 ( p_a \bar \partial \theta_a^t +
  \bar p_a \partial \bar\theta_a^t )\Bigr]  ~, \\
S_{\text{int}} \ &=\   \frac{1}{2 \pi} \int d^2 z \Bigl[\frac{1}{k} e^{2 b \phi}  +  \frac{1}{k} ( p_1 - \theta_2 ) q
 (\bar p_1 - \bar \theta_2 )^t e^{ b \phi }
  + \frac{1}{k} \bar p_2 q^{-1}  p_2^t e^{ b  \phi } \Bigr] ~, 
\end{split}
\end{equation}
where the new background charge is $Q = b^{-1} - b $.
The vertex operators are
\begin{align}
V^{ s^a_b }_{\alpha,L} (z) = e^{ i s^a_b Y^a_b + i \bar s^a_b \bar Y^a_b}
e^{2 \alpha \phi} V^\text{SU(2)}_{L} (q)~,
\end{align}
and the pre-factor is
\begin{align}
  \Theta = u \prod_{i < j}^N (z_i - z_j)^{\frac{1}{2b^2} - 1}
   \prod_{ p < q}^{N-2} (y_p - y_q)^{\frac{1}{2b^2} - 1}
   \prod_{i =1}^N \prod_{p =1}^{N-2} (z_i - y_p)^{ - \frac{1}{2b^2} + 1} ~.
\end{align}

To simplify the action we rotate the fermions as follows
\begin{equation}
\begin{split}
 \theta_1 + p_2 \ = \ \chi_- ~ ,\qquad
-\theta_2 + p_1  \ = \ \psi_- ~ ,\qquad
p_1 \ = \ \chi_+ ~ ,\qquad
p_2 \ = \ -\psi_+ ~ , \\
\bar \theta_1 + \bar p_2 \ = \ \bar\chi_- ~,\qquad
-\bar \theta_2 + \bar p_1  \ = \ \bar\psi_- ~,\qquad
\bar p_1 \ = \ \bar\chi_+ ~,\qquad
\bar p_2 \ = \ -\bar\psi_+ ~ .
\end{split}
\end{equation}
Then the action becomes
\begin{equation}\label{eq:N4action}
\begin{split}
  S\ &= \ S^{\textrm{WZNW}}[q]_{k-2} + S_0 + S_{\text{int}} ~, \\
S_0 \ &= \
\frac{1}{2 \pi} \int d^2 z \Bigl[
\bar \partial \phi \partial \phi + \frac{Q}{4} \sqrt{g} {\cal R} \phi
 + \chi_+ \bar \partial \chi_-^t + \psi_+ \bar \partial \psi_-^t
+ \bar \chi_+ \partial \bar \chi_-^t + \bar \psi_+ \partial \bar \psi_-^t
\Bigr]
~,  \\
S_{\text{int}} \ &= \ \frac{1}{2 \pi} \int d^2 z \Bigl[\frac{1}{k} e^{2 b \phi}  + \frac{1}{k}  \psi_-  q
 \bar \psi_-^t e^{ b \phi }
  +\frac{1}{k} \bar\psi_+ q^{-1} \psi_+^ t  e^{ b  \phi } \Bigr]  ~.
\end{split}
\end{equation}
Note that the fermions $\chi,\bar\chi$ have decoupled from a theory that can be called small $\mathcal N=4$ Liouville, the action of which was proposed in \cite{Ivanov:1987mz}. This has central charge
\begin{equation}\label{}
    c=c_{\textrm{SU}(2)_{k-2}}+c_{\phi}+c_{\psi}=\frac{3(k-2)}{k}+1+6Q^2+2=6(k-1) ~.
\end{equation}
In appendix \ref{symmetry} the action is shown to preserve small
$\mathcal{N}=(4,4)$ superconformal symmetry.

\subsection{From OSP($n|$2) to ${\cal N}=n$ Liouville theory}

\label{large}

We now find another family of correspondences. These are between WZNW models of orthosymplectic supergroups
OSP($n|$2) and what we call so($n$) extended super Liouville theory. For $n=1,2,3,4$ the symmetry of these theories are generated by a usual superconformal algebra,  while for higher $n$
it is a superconformal $W$-algebra.

\subsubsection{The OSP($n|$2) WZNW model}

As in the cases of OSP($n|$2) with $n=1,2$ \cite{HS2},
we parameterize the element of the supergroup as $g=\alpha G \beta$
with
\begin{equation}
\begin{split}\label{eq:paramosp}
 \alpha\ &=\ \exp ( \theta_1 F_1^-)\exp ( \theta_2 F_2^-)
 \cdots \exp ( \theta_n F_n^-)\exp(\gamma E^-)  ~,  \\
 \beta\ &=\ \exp ( \bar \theta_n F_n^+)\exp ( \bar \theta_{n-1} F_{n-1}^+)
 \cdots \exp ( \bar \theta_1 F_1^+)\exp(\bar \gamma E^+) ~, \\
 G\ &=\ e^{ - 2 \phi E^0 }
\begin{pmatrix}
  q & 0 \\
  0 & \mathbb{I}_2
\end{pmatrix} ~, \qquad q \in \text{SO($n$)} ~ .
\end{split}
\end{equation}
Notation is given in appendix \ref{notation}.
With the above parametrization
we obtain the action for OSP($n|$2) as
\begin{align}
 S^{\textrm{WZNW}}[g]_k = S^{\textrm{WZNW}}[q]_k + \frac{k}{2 \pi} \int d^2 z \Bigl[ \bar \partial \phi \partial \phi
  + e^{ - 2 \phi} (\bar \partial \gamma + \theta  \bar \partial
  \theta^t ) (\partial \bar \gamma +
     \bar \theta \partial \bar \theta^t )
 + 2 e^{ - \phi}  \bar \partial \theta  q \partial \bar \theta^t \Bigr]~,
\end{align}
where
$
\theta = (\theta_1 , \cdots , \theta_n) ,
\bar \theta = (\bar \theta_1 , \cdots , \bar \theta_n)$,
and we have used $q^t =q^{-1}$.

Introducing the auxiliary fields, $\beta, \bar \beta,
p=(p_1 , \cdots , p_n ), \bar p=(\bar p_1 , \cdots , \bar p_n )$,
we find classically
\begin{align}
 S^{\textrm{Clas.}}[g,\beta,\bar \beta,p , \bar p] = S[q]^{\textrm{WZNW}}_k + \frac{1}{2 \pi}
 \int & d^2 z \Bigl[  k \bar \partial \phi \partial \phi  -
  \beta \bar \partial \gamma - \bar \beta \partial \bar \gamma
  + p \bar \partial \theta^t  + \bar p \partial \bar \theta^t \nonumber \\
 & - \frac{1}{k} \beta \bar \beta e^{2 \phi}
  - \frac{1}{2k} (p + \beta \theta) q
  (\bar p + \bar \beta \bar \theta)^t e^{\phi} \Bigr]~.
\end{align}
Due to change of the invariant measure to the free measure, there are shifts in the coefficients. First let us
set $q=1$. Then, as in \cite{HS2}, the contribution from the measure of the path integral over
$\beta,\gamma$ is
\begin{align}
\label{anomaly1}
\delta S= - \frac{1}{\pi} \int d^2 z \partial \phi \bar \partial \phi
  + \frac{1}{8 \pi} \int d^2z  \sqrt{g} {\cal R } \phi ~,
\end{align}
and the contribution from one set of fermions $p^i, \theta_i$ is
\begin{align}
\label{anomaly2}
\delta S=  \frac{1}{4 \pi} \int d^2 z \partial \phi \bar \partial \phi
  - \frac{1}{16 \pi} \int d^2z  \sqrt{g} {\cal R } \phi ~.
\end{align}
For the $SO(n)$ part, the shift of level comes from the fermions,
and we have $k \to k - 1$.
Finally we find
\begin{align}
 S[g,\beta,\bar \beta,p , \bar p]_k = S^{\textrm{WZNW}}[q]_{k-1}  + \frac{1}{2 \pi} 
 \int & d^2 z  \Bigl[  \bar \partial \phi \partial \phi
   + \frac{\hat Q}{4}  \sqrt{g} {\cal R } \phi
- \beta \bar \partial \gamma - \bar \beta \partial \bar \gamma
  + p \bar \partial \theta^t  + \bar p \partial \bar \theta^t\nonumber \\
 & - \frac{1}{k} \beta \bar \beta e^{2 b \phi}
  - \frac{1}{2k} (p + \beta \theta) q
  (\bar p + \bar \beta \bar \theta)^t  e^{ b \phi} \Bigr]
\end{align}
with $b^{-2} = k - 2 + n/2$ and $\hat Q = b (1 - n/2)$.

As before, we bosonize the fermions as
\begin{align}
  p_a = e^{i Y_a} ~, \qquad \theta_a = e^{- i Y_a} ~, \qquad
  \bar p_a = e^{i \bar Y_a} ~, \qquad
  \bar \theta_a = e^{- i \bar Y_a}
\end{align}
with $a=1,2,\cdots,n$. Then the vertex operators are
\begin{align}
 V^{s_a}_{j,L} (\mu | z) =  \mu ^{j + 1 + \frac{1}{2} \sum_a s_a} \bar \mu ^{j + 1 + \frac{1}{2}\sum_a  \bar s_a}
 e^{ i s_a Y_a + i \bar s_a \bar Y_a }e^{\mu \gamma - \bar \mu \bar \gamma}
  e^{2b(j+1) \phi} V^\text{SO($n$)}_L (q) ~,
\end{align}
where $L$ labels the representation of so($n$).

\subsubsection{The correspondence with so($n$)-extended super Liouville theory}

We consider correlation functions of the above vertex operators and
map them to those of ${\cal N}=n$ super Liouville theories.
First $\gamma$ is integrated out, and then
 the field $\beta$ is replaced by the function
\begin{align}
 \sum_{i=1}^N \frac{\mu_i}{z-z_i} = u \frac{\prod_{l=1}^{N-2} (z - y_l)}
 {\prod_{i=1}^N (z - z_i) } = u {\cal B} (y_l , z_i ; z) ~,
\end{align}
and similarly $\bar \beta$ by $ - u \bar {\cal B} (\bar y_l , \bar z_i ; \bar z)$. Like in \eqref{eq:ShiftabsorbB} we absorb ${\cal B}, \bar {\cal B}$ by a shift of fields $\phi,Y_a$ and get
\begin{align}
 \left\langle \prod_{i=1}^N V^{ {s_a}_i }_{j_i,L_i} (\mu_i | z_i)  \right\rangle_{\textrm{WZNW}}  = \delta^{(2)} (\sum_{i=1}^N \mu_i )
 | \Theta_N |^2  \left\langle \prod_{i=1}^N
V^{ {s_a}_i +1/2 }_{b(j_i + 1) + 1/2b,L_i}  (z_i)
 \prod_{l=1}^{N-2} V^{ -1/2 }_{-1/2b,0} ( y_l)
 \right\rangle  ~.
\end{align}
Here $L=0$ is the identity representation of so($n$).
The action for the right hand side is
\begin{equation}
\begin{split}
 S\ &=\ S^{\textrm{WZNW}}[q]_{k-1} + S_0 + S_{\text{int}} ~, \\
S_0 \ &= \ \frac{1}{2 \pi} \int d^2 z \Bigl[
\bar \partial \phi \partial \phi + \frac{Q}{4} \sqrt{g} {\cal R} \phi
+ \frac{1}{2} \bigl( p \bar \partial \theta^t - \bar\partial p \theta^t + \bar p \partial \bar \theta^t- \partial \bar p \bar\theta^t\bigr)\Bigr] ~, \\
S_{\text{int}} \ &=\  \frac{1}{2 \pi} \int d^2 z \Bigl[ \frac{1}{k} e^{2 b \phi}  - \frac{1}{2 k} ( p + \theta ) q
 (\bar p - \bar \theta )^t e^{ b \phi } \Bigr]
\end{split}
\end{equation}
with $Q = \hat Q + b^{-1}$.
The vertex operators are
\begin{align}
V^{ s_a }_{\alpha,L} (z) = e^{ i s_a Y_a + i \bar s_a \bar Y_a}
e^{2 \alpha \phi} V^\text{SO($n$)}_{L} (q) ( z) ~,
\end{align}
and the pre-factor is
\begin{align}
  \Theta = u \prod_{i < j}^N (z_i - z_j)^{\frac{1}{2b^2} - \frac{n}{4}}
   \prod_{ p < q}^{N-2} (y_p - y_q)^{\frac{1}{2b^2} - \frac{n}{4}}
   \prod_{i =1}^N \prod_{p =1}^{N-2} (z_i - y_p)^{ - \frac{1}{2b^2} + \frac{n}{4}} ~.
\end{align}

If we rotate fermions as
\begin{align}
 p  = \frac{1}{\sqrt{2}} (\chi + i \psi) ~,  \qquad
 \theta = - \frac{1}{\sqrt{2}} (\chi - i \psi) ~, \qquad
\bar p = \frac{1}{\sqrt{2}} (\bar \chi + i \bar \psi) ~,  \qquad
\bar \theta =  \frac{1}{\sqrt{2}} (\bar \chi - i \bar \psi) ~,
\nonumber
\end{align}
then the fermions $\chi$ decouple from
\begin{align} \label{eq:slft}
 S^\text{SL}[\phi,\psi,q] =
S[q]_{k-n/2} + \frac{1}{2 \pi} \int d^2 z \Bigl[
\bar \partial \phi \partial \phi &+ \frac{Q}{4} \sqrt{g} {\cal R} \phi
+ \frac{1}{2} \psi \bar \partial \psi^t +  \frac{1}{2} \bar \psi \partial \bar \psi^t\nonumber \\
  &+ \frac{1}{k} e^{2 b \phi}  + \frac{1}{ k} \psi q  \bar \psi^t e^{ b \phi } \Bigr] ~,
\end{align}
which may be named so($n$)-extended super Liouville field theory.
The central charge is:
\begin{align}\label{}
    c_{\textrm{Liouville}}&=c_{\mathrm{SO}(n)}+c_{\psi}+c_\phi=\frac{(n-1)(n-2)}{2}\frac{k-1}{k+n/2-2}+n/2+1+6 Q^2\nonumber\\
    &=\tfrac{1}{2}S\frac{6S+n^2-10}{S+n-3},
\end{align}
where $S=2k-1$. In particular, the actions for $n=3,4$ were proposed
in \cite{Ivanov:1987mz}. For $n\leq3$ this has $\mathcal N=(n,n)$ superconformal symmetry. For $n=4$ we get the large $\mathcal N=(4,4)$ superconformal algebra, and for $n>4$ we get Knizhnik's so($n$)-extended superconformal algebra \cite{K}, see appendix \ref{symmetry}.

For the vertex operators we can factor out the component corresponding to the decoupled fermions $\chi$. This is done just like the OSP(1$|$2) case in \cite{HS2} by introducing spin fields for the fermions $\psi,\chi$.

\section{From supergroups to superconformal $W$-algebras}

\label{general}

We now generalize the correspondence. Our derivation works for any Lie supergroup whose bosonic part is
of the form $H_3^+\times A$ for some Lie group $A$.
We first introduce the relevant Lie superalgebras, and then write down the WZNW model using
a suitable parametrization of the supergroup valued field. The derivation of the correspondences is analogous to the previous section.
The superconformal $W$-algebras are introduced in appendix \ref{ESA} using the works
\cite{IM,Ito:1992bi,Ito:1993qi}.

\subsection{Lie superalgebras}

Our goal is to find and derive correspondences between supergroup WZNW models and conformal field theories with
some superconformal $W$-algebra symmetry. The supergroups have to contain SL(2,$\mathbb R$) as a factor of the bosonic subgroup (i.e. $H^3_+$ after a coset construction). Moreover it has to be possible to write the action of the model in a suitable form that allows for integration of the fields $\gamma, \bar\gamma$.
Finding this suitable action is equivalent to finding a certain grading for the superalgebra. This grading is as follows.
Let $\g$ be a simple superalgebra whose bosonic subalgebra is  sl(2) $\oplus$ $\sub$ for some reductive Lie algebra $\sub$.
Denote by $\pm\nu$ the positive and negative root of sl(2) with corresponding generators $E_{\pm\nu}$. These commute to the Cartan element $H_\nu$ and we have the commutation relations
\begin{align}\label{}
    [E_\nu,E_{-\nu}]=2H_\nu\ ,\qquad [H_\nu,E_{\pm\nu}]=\pm E_{\pm \nu}\ .
\end{align}
We will see that there is a correspondence for WZNW models whose algebra allows a five-decomposition of the form (see also \cite{Kac:2003jh})
\begin{equation}
\g \ = \ \g_{-1}\oplus\g_{-1/2}\oplus\g_0\oplus\g_{1/2}\oplus\g_1\, ,
\end{equation}
where the spaces $g_i$ are bosonic for integer $i$ and otherwise fermionic. The index labels the eigenvalue of the adjoint action of $H_\nu$, so $\g_{\pm 1}=\text{span}\{E_{\pm\nu}\}$, and $\g_0=\text{span}\{H_\nu\}\oplus\sub$. Let $\Delta^1_\pm$ denote the roots corresponding to $\g_{\pm 1/2}$, then $\Delta^1_-+\nu=\Delta^1_+$. We will always normalize our bilinear form $\kappa$ such that $\nu$ has negative length, and the
longest positive root of positive norm has norm two (this will then lie in $\sub$). We can then characterize $\Delta^1_+$ as
\begin{equation}
\Delta_+^1 \ = \ \{ \gamma \,\in\,\Delta^1 \, | \, (\gamma, \nu) < 0\,\}\, ,
\end{equation}
and oppositely is $\Delta^1_-$ positive with respect to $\nu$.

We now give examples of such Lie superalgebras. Looking back at \eqref{eq:parampsl22} and \eqref{eq:paramosp} we already had psl(2$|$2) and osp($n|$2) as examples. We now list further interesting examples for classical Lie super algebras, as reference see \cite{Dictionary}.

\subsection{Examples}

We now give examples of the possible Lie superalgebras with sl(2) $\oplus$ $\sub$ as bosonic Lie algebra and a five-decomposition. As above we always have $\g_{\pm 1}=\text{span}\{E_{\pm\nu}\}$, and $\g_0=\text{span}\{H_\nu\}\oplus\sub$.

\subsubsection*{sl($n|$2)}

The rank of sl($n|$2) is $n+1$, the dimension $n^2+n+1$ and the bosonic subalgebra is
\begin{equation}
 \text{sl}(n) \oplus \text{u}(1)\oplus \text{sl}(2)
\ = \ \text{sl}(2) \oplus \sub\, .
\end{equation}
The root lattice is generated by vectors $\epsilon_1,...,\epsilon_n$ and $\delta_1,\delta_2$ with non-zero inner products
$\epsilon_i\epsilon_i=1$ and $\delta_i\delta_i=-1$.
The root spaces are
\begin{equation}
 \begin{split}
\Delta_0 \ &= \ \{ \epsilon_i-\epsilon_j,\delta_i-\delta_j\} ~,\qquad
  \Delta_1 \ = \ \{\pm(\epsilon_i-\delta_j)\} ~ .
 \end{split}
\end{equation}
The sl(2) root is $\nu=\delta_1-\delta_2$ and its norm is $\nu\nu=-2$.
The longest root of $\sub$ has norm $\alpha_L^2/2=1$.
We decompose sl($n|$2) as follows:
\begin{equation}
\begin{split}
 \text{sl}(n|2) \ &= \  \g_{-1}\oplus\g_{-1/2}\oplus\g_0\oplus\g_{1/2}\oplus\g_1 ~, \\
\g_{-1/2}\ &= \ \text{span}\{ F_{-\epsilon_i+\delta_2}, F_{\epsilon_i-\delta_1}\} ~, \\
\g_{1/2} \ &= \ \text{span}\{ F_{\epsilon_i-\delta_2}, F_{-\epsilon_i+\delta_1}\} ~ .
\end{split}
\end{equation}

\subsubsection*{F(4)}

The exceptional Lie superalgebra F(4) has rank 4, dimension 40, and its bosonic subalgebra is
\begin{equation}
 \text{sl}(2)\oplus \text{so}(7) \ = \ \text{sl}(2)\oplus\sub\, .
\end{equation}
The root lattice is generated by vectors $\delta, \epsilon_1,\epsilon_2,\epsilon_3$ with non-zero inner product $\epsilon_i\epsilon_i=1$
and $\delta\delta=-3$.
The root spaces are
\begin{equation}
 \Delta_0\ = \ \{\pm\delta,\pm\epsilon_i\pm\epsilon_j,\pm\epsilon_i \, |\, i\neq j\} ~ ,\qquad
\Delta_1\ = \ \{\tfrac{1}{2}(\pm\delta\pm \epsilon_1\pm\epsilon_2\pm\epsilon_3)\} ~ .
\end{equation}
We denote the generators for the sl(2) roots by $E_{\pm\nu}=E_{\pm\delta}$.
Note, that $\nu\nu=\delta\delta=-3$. The longest root of $\sub$ has norm $\alpha_L^2/2=1$.
Then the decomposition is
\begin{equation}
\begin{split}
F(4)\ &= \ \g_{-1}\oplus\g_{-1/2}\oplus\g_0\oplus\g_{1/2}\oplus\g_1 ~, \\
\g_{-1/2}\ &= \ \text{span}\{ F_{\frac{1}{2}(-\delta\pm\epsilon_1\pm\epsilon_2\pm\epsilon_3)}\} ~, \\
\g_{1/2}\ &= \ \text{span}\{ F_{\frac{1}{2}(+\delta\pm\epsilon_1\pm\epsilon_2\pm\epsilon_3)}\} ~ .
\end{split}
\end{equation}

\subsubsection*{G(3)}

The exceptional Lie superalgebra G(3) has rank 3, dimension 31, and its bosonic subalgebra is
\begin{equation}
 \text{sl}(2)\oplus \text{g}_2 \ = \ \text{sl}(2)\oplus\sub\, .
\end{equation}
The root lattice is generated by vectors $\delta, \epsilon_1,\epsilon_2,\epsilon_3$
with non-zero inner product
\begin{equation}
 \epsilon_i\epsilon_j\ =\ -\frac{1}{3}+\delta_{ij} ~,\qquad \delta\delta \ = \ -\frac{2}{3}\, .
\end{equation}
The root spaces are
\begin{equation}
 \Delta_0\ = \ \{\pm2\delta,\epsilon_i-\epsilon_j,\pm\epsilon_i \,|\,i\neq j \} ~,\qquad
\Delta_1\ = \ \{\pm\delta\pm \epsilon_i, \pm\delta\}\, .
\end{equation}
We denote the generators for the sl(2) roots by $E_{\pm\nu}=E_{\pm2\delta}$.
Note, that $\nu\nu=4\delta\delta=-8/3$. The longest root of $\sub$ has norm $\alpha_L^2/2=1$
Then the decomposition is
\begin{equation}
\begin{split}
G(3)\ &= \ \g_{-1}\oplus\g_{-1/2}\oplus\g_0\oplus\g_{1/2}\oplus\g_1 ~, \\
\g_{-1/2}\ &= \ \text{span}\{  F_{-\delta\pm\epsilon_i}, F_{-\delta}\} ~, \\
\g_{1/2}\ &= \ \text{span}\{  F_{+\delta\pm\epsilon_i}, F_{\delta}\} ~ .
\end{split}
\end{equation}

\subsubsection*{D(2,1;$\alpha$)}
There is a family of  exceptional Lie superalgebras D(2,1;$\alpha$) parameterized by $\alpha\neq0,-1,\infty$. There are a self-dualities taking $\alpha\mapsto 1/\alpha$ and $\alpha\mapsto -1-\alpha$, and we use these to choose $\alpha\in \left]-1,0\right[$.
The superalgebra has rank 3, dimension 17, and its bosonic subalgebra is
\begin{equation}
 \text{sl}(2)\oplus \text{sl}(2)\oplus \text{sl}(2) \ = \ \text{sl}(2)\oplus\sub\, .
\end{equation}
The root lattice is generated by vectors $\epsilon_1,\epsilon_2,\epsilon_3$
with non-zero inner product
\begin{equation}
 \epsilon_1\epsilon_1\ =\ -\frac{(1+\alpha)}{2} ~,\qquad
\epsilon_2\epsilon_2\ =\ \frac{1}{2} ~,\qquad
\epsilon_3\epsilon_3\ =\ \frac{\alpha}{2} ~.
\end{equation}
The root spaces are
\begin{equation}
 \Delta_0\ = \ \{\pm2\epsilon_i \} ~,\qquad
\Delta_1\ = \ \{\pm\epsilon_1\pm\epsilon_2\pm\epsilon_3\} ~ .
\end{equation}
We denote the generators for the sl(2) roots we want to reduce by $E_{\pm\nu}=E_{\pm2\epsilon_1}$.
Note, that $\nu\nu=4\epsilon_1\epsilon_1=-2-2\alpha$. The longest root of $\sub$ have norm $\alpha_L^2/2=\alpha$
for one sl(2) and the other one has $\alpha_L^2/2=1$.
Then the decomposition is
\begin{equation}
\begin{split}
D(2,1;\alpha)\ &= \ \g_{-1}\oplus\g_{-1/2}\oplus\g_0\oplus\g_{1/2}\oplus\g_1 ~, \\
\g_{-1/2}\ &= \ \text{span}\{  F_{-\epsilon_1\pm\epsilon_2\pm\epsilon_3}\} ~, \\
\g_{1/2}\ &= \ \text{span}\{   F_{\epsilon_1\pm\epsilon_2\pm\epsilon_3}\} ~ .
\end{split}
\end{equation}

\subsubsection*{spo($2n|4$)}

The rank of spo($2n|4$) is $n+2$, the dimension $2n^2+9n+6$ and the bosonic subalgebra is
\begin{equation}
 \text{sp}(2n) \oplus \text{so}(4) \ = \ \text{sl}(2) \oplus \sub\, .
\end{equation}
The root lattice is generated by vectors $\delta_1,...,\delta_n$ and $\epsilon_1,\epsilon_2$ with non-zero inner products
$\delta_i\delta_i=1/2$ and $\epsilon_i\epsilon_i=-1/2$.
The root spaces are
\begin{equation}
 \begin{split}
\Delta_0 \ &= \ \{ \pm\epsilon_1\pm\epsilon_2,\pm\delta_i\pm\delta_j,\pm2\delta_n\,|\, i\neq j\} ~,\qquad
  \Delta_1 \ = \ \{\pm\epsilon_i\pm\delta_j\} ~ .
 \end{split}
\end{equation}
The sl(2) root is $\nu=\epsilon_1+\epsilon_2$ and its norm is $\nu\nu=-1$.
The longest root of $\sub$ has norm $\alpha_L^2/2=-1/2$ for the sl(2) and $\alpha_L^2/2=1$ for the sp$(2n)$.
Then the decomposition is
\begin{equation}
\begin{split}
\text{spo}(2n|4)\ &= \ \g_{-1}\oplus\g_{-1/2}\oplus\g_0\oplus\g_{1/2}\oplus\g_1 ~, \\
\g_{-1/2}\ &= \ \text{span}\{ F_{-\epsilon_i\pm\delta_a}\} ~, \\
\g_{1/2}\ &= \ \text{span}\{ F_{\epsilon_i\pm\delta_a}\} ~ .
\end{split}
\end{equation}

In table \ref{table:examples} we summarize all the examples with some data needed in the next subsection. Let us also note that spo($2n|3)$  has bosonic subgroup sl(2)$\oplus$sp(2$n)$, but there is no five-decomposition of our wanted form and hence it is not included.

\begin{table}
\begin{center}
 \begin{tabular}{ |c| c | c | c | c | c | c | c |}
    \hline
    $\g$ & $\sub$ & $h^\vee$ $\vphantom{\Bigl(\Bigr)}$ & $ h^\vee_{\sub}$ & $\nu^2$ & $N_f$  &$\alpha^2_L/2$\\[1mm] \hline\hline
 psl(2$|$2) & sl(2) & $0$ & $2$ & $-2$ & $4$  & $1$ \\  \hline
    osp($n|$2) \  & so($n$) & $n-4$ & $n-2$ & $-4$ & $n$  & $1$ \\  \hline
    sl($n|2$)\ , $n\neq 2$ & sl($n$)$\oplus$ u(1)  & $n-2$ & $(n,0)$ & $-2$ & $2n$  & $1$ \\  \hline
 F(4) & so(7) & $3$ & $5$ & $-3$ & $8$  & $1$ \\  \hline
    G(3) & g$_2$ & $2$ & $4$ & $-8/3$ & $7$  & $1$ \\  \hline
 D(2,1;$\alpha$)& sl(2)$\oplus$sl(2) & $0$ & $(2,2)$ & $-2-2\alpha$ & $4$  & $(1,\alpha) $ \\  \hline
   spo($2n|4$)  & sl(2)$\oplus$sp($2n$) & $n-1$ & $(2,n+1)$ & $-1$ & $4n$  & $(-1/2,1)$ \\  \hline
\end{tabular}\caption{{\em Summary of the examples with five-decomposition along with useful constants. $h^\vee$ is the dual Coxeter number of $\g$, $h^\vee_{\sub}$ are the dual Coxeter numbers of the simple parts of $\sub$, $N_f$ is half the number of fermionic generators, and $\alpha^2_L$ are the lengths of the longest roots in the simple parts of $\sub$. Compared to  last section, the psl(2$|$2) and osp($\mathit{n|2})$ cases have been normalized standardly with the longest positive root of positive norm having length two (in last section the bilinear forms used where respectively minus supertrace and supertrace, for osp($\mathit{n|2}$) the standard bilinear form would have been half the supertrace).}}\label{table:examples}
\end{center}
\end{table}

\subsection{The supergroup WZNW model}

We can parameterize the group valued fields
as
\begin{align}\label{}
    g\ =\ g_{-1}\, g_{-1/2}\, g_0\,  g_{1/2}\, g_1,
\end{align}
and introduce bosonic fields $\gamma$ lying in $\g_{-1}=\text{span}\{E_{-\nu}\}$, $\bar\gamma$ in $\g_1=\text{span}\{E_{\nu}\}$, and fermionic fields $\theta$ in $\g_{-1/2}$ and $\bar\theta$ in $\g_{1/2}$ such that
\begin{align}\label{}
    g_{-1}&=e^{\gamma}\ , & g_{-1/2}&=e^{\theta}\ , &
    g_{1/2}&=e^{\bar\theta}\ , &g_{1}&=e^{\bar\gamma}\ .
\end{align}
Then the action is of the form
\begin{equation}
S^{\textrm{WZNW}}[g]_k\ = \ S^{\textrm{WZNW}}[g_0]_k + \frac{k}{2\pi}\int d^2z\,  \langle (\bar\del\gamma+\tfrac{1}{2}[\bar\del\theta, \theta]+\bar\del\theta),\Ad(g_0)
 (\del\bar\gamma-\tfrac{1}{2}[\del\bar\theta, \bar\theta]+\del\bar\theta)\rangle \, .
\end{equation}
 Using the five-decomposition we know that $ \Ad (g_0)$ leaves $\g_i$ invariant and $[\del\bar\theta, \bar\theta]$ lies in $\text{span}\{E_{\nu}\}$. Thus we can introduce auxiliary fields $\beta$ taking values in $\text{span}\{E_{\nu}\}$ and $\bar\beta$ taking values in $\text{span}\{E_{-\nu}\}$ and get the classical equivalent action
\begin{align}\nonumber
S^{\textrm{Clas.}}[g,\beta,\bar\beta]_k\ &= \ S^{\textrm{WZNW}}[g_0]_k + \frac{k}{2\pi}\int d^2z\,
\langle \beta, (\bar\del\gamma+\tfrac{1}{2}[\bar\del\theta,\theta])\rangle +
\langle \bar\beta, (\del\bar\gamma-\tfrac{1}{2}[\del\bar\theta,\bar\theta])\rangle + \\
&\qquad\qquad\qquad\qquad\qquad
+\langle \bar\del\theta, \Ad(g_0)\del\bar\theta\rangle -\langle\beta, \Ad(g_0)\bar\beta\rangle \,.
\end{align}
Further, we introduce fermionic auxiliary fields $p,\bar p$ taking values in $\g_{\pm 1}$ and classically we get (here we need to use the invariance of the bilinear form)
\begin{align}\nonumber
S^{\textrm{Clas.}}[g,\beta,\bar\beta, p,\bar p]_k\ &= \ S^{\textrm{WZNW}}[g_0]_k + S_0[\beta,\bar\beta,\gamma, \bar\gamma, p,\bar p, \theta, \bar\theta] + S_{\text{int}}[\beta,\bar\beta, \theta, \bar\theta, p, \bar p, g_0] ~, \\
S_0 \ &= \ \frac{k}{2\pi}\int d^2z\, \langle \beta ,\bar\del\gamma\rangle + \langle \bar\beta , \del\bar\gamma\rangle + \langle p , \bar\del\theta \rangle + \langle \bar p , \del\bar \theta\rangle ~, \\
 S_{\text{int}} \ &= \ -\frac{k}{2\pi}\int d^2z\, \langle \beta, \Ad(g_0)\bar\beta\rangle + \langle (p+\tfrac{1}{2}[\beta, \theta]), \Ad(g_0)(\bar p-\tfrac{1}{2}[\bar\beta, \bar\theta])\rangle \, .
\nonumber
\end{align}
Now, we parameterize $g_0= q e^{-2\phi H_\nu}$ with $H_\nu$ being the Cartan direction of the sl(2). Then the action becomes
\begin{equation}
\begin{split}
S^{\textrm{Clas.}}[g,\beta,\bar\beta, p,\bar p]_k\, = \, S^{\textrm{WZNW}}[q]_k + S[\phi]+ 
S_0[\beta,\bar\beta,\gamma, \bar\gamma, p,\bar p, \theta, \bar\theta]  + S_{\text{int}}[\beta,\bar\beta, \theta, \bar\theta, p, \bar p, q,\phi] ~, \\
\end{split}
\end{equation}
with
\begin{equation}
\begin{split}
S[\phi] \ &= \ -\frac{k}{\nu^2\pi}\int d^2z\, \langle \del\phi , \bar\del\phi\rangle ~, \\
S_0 \ &= \ \frac{k}{2\pi}\int d^2z\, \langle \beta , \bar\del\gamma\rangle + \langle \bar\beta , \del\bar\gamma\rangle + \langle p , \bar\del\theta\rangle + \langle \bar p , \del\bar \theta\rangle ~, \\
 S_{\text{int}} \ &= \ -\frac{k}{2\pi}\int d^2z\, e^{2\phi}\langle \beta, \bar\beta\rangle
 + e^{\phi}\langle (p+\tfrac{1}{2}[\beta, \theta]), \Ad(q)(\bar p-\tfrac{1}{2}[\bar\beta, \bar\theta])\rangle \, ,
\end{split}
\end{equation}
where we used that the grading indices in $\g_i$ simply are the eigenvalues of $\Ad(H_\nu)$ and $\left<H_\nu,H_\nu\right>=1/\nu^2$.

Let us now take into account the quantum effects. First we consider the change in levels. Using  \eqref{anomaly1} and \eqref{anomaly2} we see that the level for the $\phi$-part changes as
\begin{align}\label{}
    k\mapsto k-\frac{\nu^2}{2}(-2+\frac{N_f}{2})=k+h^\vee.
\end{align}
Here $N_f=|\Delta^1_+|$ is the number of fermions $p$. As we have also written, the change in the level can be seen to be the dual Coxeter number of the superalgebra, i.e. half the eigenvalue of the quadratic Casimir in the adjoint representation. Likewise the level for a simple component, $\sub_i$, of $\sub$ change with half the eigenvalue of the Casimir, $C_{\textrm{free}}$ in the representation that the free fermions transform in. To calculate this we note that quadratic Casimir in the adjoint representation splits into the a quadratic adjoint Casimir for the $\sub_i$ component and $C_{\textrm{free}}$
\begin{equation}
C_{\text{ad}} \ = \ C_{\sub_i} + C_{\text{free}}\, .
\end{equation}
However, $C_{\sub_i}$ is not canonically normalized. Our bilinear form, $\langle\ ,\ \rangle$, for the Lie superalgebra is canonically normalized such that the
longest positive root with positive norm has norm two \cite{Kac:1994kn}. This means, we have a relation to the canonically normalized bilinear form, $\langle\ ,\ \rangle_{\sub_i}$, for the subalgebra as
\begin{equation}
 \langle\ \ , \ \ \rangle_{\sub_i} \ = \ \frac{2}{\alpha_L^2} \langle\ \ , \ \ \rangle|_{\sub_i}\, ,
\end{equation}
where $\alpha^2_L$ is the length of the longest root of $\sub_i$ measured by $\langle\ ,\ \rangle$. This means that half the eigenvalue of $C_{\sub_i}$ is $\frac{\alpha_L^2}{2}h^\vee_{\sub_i}$, where $h^\vee_{\sub_i}$ is the dual Coxeter number for $\sub_i$. The renormalized level for the WZNW model on the $\sub_i$ component with the standard bilinear form $\langle\ ,\ \rangle_{\sub_i}$ is thus given by the formula
\begin{equation}
 k_{\sub_i} \ = \ \frac{2}{\alpha_L^2}\bigl(k+h^\vee-\frac{\alpha_L^2}{2}h^\vee_{\sub_i}\bigr)\, .
\end{equation}
A similar formula holds for the u(1) part in the psl($n|2$) case if we do not renormalize, i.e. its level is simply changed by the dual Coxeter number.

Finally, also background charges for the field $\phi$ appear as seen from
\eqref{anomaly1} and \eqref{anomaly2}. Rescaling $\beta,\bar\beta, p, \bar p$ and $\phi$ the quantum corrected action then becomes
\begin{align}
S[g,\beta,\bar\beta, p,\bar p]\ &= \ S^{\textrm{WZNW}}[q] + S[\phi]+ S_0[\beta,\bar\beta,\gamma, \bar\gamma, p,\bar p, \theta, \bar\theta] + S_{\text{int}}[\beta,\bar\beta, \theta, \bar\theta, p, \bar p, q,\phi] ~, \nonumber \\
S^{\textrm{WZNW}}[q] \ &= \ \sum_i S^{\textrm{WZNW}}[q_i]_{k_{\sub_i}} ~, \nonumber\\
S[\phi] \ &= \ \frac{1}{2\pi}\int d^2z\,  \del\phi\bar\del\phi + \frac{\hat Q}{4}\sqrt{g}\mathcal R \phi ~, \\
S_0 \ &= \ \frac{1}{2\pi}\int d^2z\, \langle \beta , \bar\del\gamma\rangle + \langle \bar\beta , \del\bar\gamma\rangle + \langle p , \bar\del\theta\rangle + \langle \bar p , \del\bar \theta\rangle ~, \nonumber\\
 S_{\text{int}} \ &= \ -\frac{1}{2k\pi}\int d^2z\, e^{2b\phi}\langle \beta, \bar\beta\rangle
 + e^{b\phi}\langle (p+\tfrac{1}{2}[\beta, \theta]), \Ad(q)(\bar p-\tfrac{1}{2}[\bar\beta, \bar\theta])\rangle ~ ,\nonumber
\end{align}
Where the sum in $S^{\textrm{WZNW}}[q]$ is over the simple and u(1) parts of $\sub$, and $q=\prod_iq_i$ is a factorization into these parts. Further
\begin{equation}
 b \ = \ \sqrt{\frac{-\nu^2}{2(k+h^\vee)}} ~ ,\qquad \hat Q \ = \ b(1-N_f/2) ~ ,\qquad N_f\ = \ |\Delta_1^+|\, .
\end{equation}
The relevant constants used can be found in table \ref{table:examples}.

\subsection{The correspondence with extended superconformal algebras}

We consider correlation functions of supergroup WZNW model and
map them to those of Liouville theories with extended
superconformal algebra. Let us introduce a basis $t^a$ for $\g_{1/2}$ and $t_a$ for $\g_{-1/2}$ with $\langle t^a,t_b\rangle=\delta^a_b$ and $a=1,2,\cdots,N_f$. We parameterize the fermions as $p=p_a t^a$ etc. and bosonize as
\begin{align}
  p_a = e^{i Y_a} ~, \qquad \theta^a = e^{- i Y_a} ~, \qquad
  \bar p^a = e^{i \bar Y_a} ~, \qquad
  \bar \theta_a = e^{- i \bar Y_a}\ .
\end{align}
For the bosonic fields let $\gamma=\gamma_{-\nu} E_{-\nu}$, $\beta=\beta_\nu E_\nu$, etc. The vertex operators of the supergroup WZNW model are then written as
\begin{align}
 V^{s_a}_{j,L} (\mu | z) =  \mu ^{j + 1 + \frac{1}{2} \sum_a s_a} \bar \mu ^{j + 1 + \frac{1}{2}\sum_a  \bar s_a}
 e^{ i s_a Y_a + i \bar s_a \bar Y_a }e^{\mu \gamma_{-\nu} - \bar \mu \bar \gamma_\nu}
  e^{2b(j+1) \phi} V^\sub_L (q) ~,
\end{align}
where $L$ labels the representation of $\sub$.

First, $\gamma_{-\nu}$ is integrated out, then
 the field $\beta_\nu$ is replaced by the function (using $\langle E_{-\nu},E_\nu\rangle=2/\nu^2$)
\begin{align}
 \beta_\nu\mapsto\frac{\nu^2}{2}\sum_{i=1}^N \frac{\mu_i}{z-z_i} = u \frac{\prod_{l=1}^{N-2} (z - y_l)}
 {\prod_{i=1}^N (z - z_i) } = u {\cal B} (y_l , z_i ; z) ~,
\end{align}
and similarly $\bar \beta_{-\nu}$ by $ - u \bar {\cal B} (\bar y_l , \bar z_i ; \bar z)$. Shifting the fields $\phi$ and $Y_a$ to absorb this function we get
\begin{align}
 \left\langle \prod_{i=1}^N V^{ {s_a}_i }_{j_i,L_i} (\mu_i | z_i)  \right\rangle_{\textrm{WZNW}}  = \delta^{(2)} (\sum_{i=1}^N \mu_i )
 | \Theta_N |^2  \left\langle \prod_{i=1}^N
V^{ {s_a}_i +1/2 }_{b(j_i + 1) + 1/2b,L_i}  (z_i)
 \prod_{l=1}^{N-2} V^{ -1/2 }_{-1/2b,0} ( y_l)
 \right\rangle  ~.
\end{align}
Here $L=0$ is the identity representation of $\sub$.
The action for the right hand side is
\begin{equation}
\begin{split}
S\ &= \ S^{\textrm{WZNW}}[q] + S[\phi]+ S_0[ p,\bar p, \theta, \bar\theta] + S_{\text{int}}[ p, \bar p, \theta, \bar\theta, q,\phi] ~, \\
S^{\textrm{WZNW}}[q] \ &= \ \sum_i S^{\textrm{WZNW}}[q_i]_{k_{\sub_i}} ~, \\
S[\phi] \ &= \ \frac{1}{2\pi}\int d^2z\,  \del\phi\bar\del\phi + \frac{ Q}{4}\sqrt{g}\mathcal R \phi ~, \\
S_0 \ &= \ \frac{1}{4\pi}\int d^2z\, \langle p , \bar\del\theta\rangle - \langle \theta , \bar\del p\rangle +
\langle \bar p , \del\bar \theta\rangle -\langle \bar\theta , \del\bar p\rangle ~, \\
 S_{\text{int}} \ &= \ -\frac{1}{2k\pi}\int d^2z\, e^{2b\phi}\frac{2}{\nu^2}
 + e^{b\phi}\langle (p+\tfrac{1}{2}[E_\nu, \theta]), \Ad(q)(\bar p-\tfrac{1}{2}[E_{-\nu}, \bar\theta])\rangle ~. 
\end{split}
\end{equation}
Recall that $E_\nu$ maps $\g_{-1/2}$ to $\g_{1/2}$.
The new background charge is $Q = \hat Q + b^{-1}$.
The vertex operators are
\begin{align}
V^{ s_a }_{\alpha,L} (z) = e^{ i s_a Y_a + i \bar s_a \bar Y_a}
e^{2 \alpha \phi} V^\sub_{L} (q) ( z) ~,
\end{align}
and the pre-factor is
\begin{align}
  \Theta = u \prod_{i < j}^N (z_i - z_j)^{\frac{1}{2b^2} - \frac{N_f}{4}}
   \prod_{ p < q}^{N-2} (y_p - y_q)^{\frac{1}{2b^2} - \frac{N_f}{4}}
   \prod_{i =1}^N \prod_{p =1}^{N-2} (z_i - y_p)^{ - \frac{1}{2b^2} + \frac{N_f}{4}} ~.
\end{align}

We may rotate the fermions as
\begin{align}
 \psi &=  p + \frac{1}{2}[E_\nu,\theta]\ , &
\bar \psi  &=  \bar p - \frac{1}{2}[E_{-\nu},\bar\theta]\ , \nonumber\\
 \chi &=  p - \frac{1}{2}[E_\nu,\theta]\ , &
\bar \chi  &=  \bar p + \frac{1}{2}[E_{-\nu},\bar\theta]\ ,
\end{align}
and we see that the fermions $\chi$ decouple from
\begin{equation}
\begin{split}\label{eq:generalsuperliouville}
 S^{\textrm{sL}}[\phi,\psi,q] &=
S[q]_{k_0} + \frac{1}{2 \pi} \int d^2 z \Bigl[
\bar \partial \phi \partial \phi + \frac{Q}{4} \sqrt{g} {\cal R} \phi
+ \langle \psi,  [E_{-\nu},\bar\partial\psi]\rangle  + \\
&\qquad\qquad\qquad\qquad + \langle \bar \psi , [E_{\nu},\partial \bar \psi]\rangle
    + e^{b\phi}\langle \psi, \Ad(q)\bar \psi\rangle \Bigr] ~,
\end{split}
\end{equation}
which we call $\sub$-extended super Liouville field theory. Note that we have dropped the contact term in the interaction.
We explain the symmetry of these theories in appendix \ref{symmetry}.

\section{Discussions}

In this note, we have studied relations between supergroup
WZNW models and Liouville theories with extended superconformal algebras.
We have shown that
$N$-point functions of tachyon vertex operators in supergroup WZNW
models on spheres can be written in terms of $(2N-2)$-point functions
in Liouville-like theories with $N-2$ extra insertions.
The relation was studied in \cite{HS2} for OSP$(n|2)$, $n=1,2$,
and we have extended the analysis to more generic cases.
First, we studied explicitly the two specific cases of PSU$(1,1|2)$ and OSP$(n|2)$
which are the most interesting models for applications in superstring theory.
Then we developed a more abstract formalism to examine general cases with
supergroups whose bosonic subgroup is of the form SL(2)$ \times A$.
Relations between these two types of theories were given using quantum Hamiltonian
reduction in \cite{BO,IM,Ito:1992bi,Ito:1993qi}. However, here we
have given direct relation of correlators and would like to stress that our relation is different from the Hamiltonian
reduction, among others there is no restriction of the momentum space in the supergroup
WZNW models.

There are several problems to consider in the future
addition to the generalization of the FZZ duality \cite{FZZ,Hikida:2008pe,Creutzig:2010bt}
mentioned in the introduction.
First of all we would like to understand
more about the generic cases. One of the important examples is
D(2,1;$\alpha$) since in this case the Liouville field theory admits
a one parameter family of ${\cal N}=(4,4)$ superconformal symmetry
\cite{Schoutens:1988ig,STP,IKL}.
We have also studied exceptional cases G(3) and F(4), and
it would be interesting to find applications in superstring theory.
In this note we have considered only amplitudes on a sphere. This extends straightforwardly to
the case with generic Riemann surfaces of higher
genus following \cite{HS}, and it would be worthwhile also to consider disk amplitudes \cite{Hosomichi:2006pz,Fateev:2007wk,Creutzig:2010zp}. Moreover, the Liouville theories with extended superconformal $W$-algebras, that we have arrived at, should be studied. We have given the precise actions of these theories and a next important step is to find their possible dualities. 
This is because in particular these in turn can be used to derive dualities for the original theories and cosets thereof. Using the explicit form of actions, it might be also possible to compute
correlation functions of these theories.

Recently, it was proposed in \cite{Alday:2009aq,Wyllard:2009hg} that two
dimensional Toda theory is related to four dimensional SU($N$) gauge theory.
Moreover, including surface operators in the gauge theory is argued to change
the two dimensional theory to the one with the symmetry of current algebra
or $W$-algebra \cite{Alday:2010vg,Kozcaz:2010yp,Wyllard:2010rp,Wyllard:2010vi}.
One family of $W$-algebras that appears in this relation are the Bershadsky-Polyakov algebras $W_N^{N-1}$ \cite{Wyllard:2010vi}, where
these algebras can be constructed from SL($N$) WZNW models via Drindfel'd-Sokolov reduction \cite{Bershadsky:1990bg}.
The construction uses a so-called good minimal graduation of sl($N$). But this is exactly the type of decomposition we needed for
our superalgebras.
Hence, it is reasonable to expect that our derivation carries over to a correspondence between SL($N$) WZNW models and
theories with $W_N^{N-1}$ symmetry. We will report on this in a forthcoming paper \cite{CHR3}.

\subsection*{Acknowledgement}

We would like to thank Louise Dolan, Thomas Quella and Volker Schomerus for useful
discussions.
The work of YH is supported in part by Keio Gijuku Academic Development Funds, and the work of TC partially by U.S. Department of Energy,
Grant No. DE-FG02-06ER-4141801, Task A.

\appendix

\section{Notations}
\label{notation}
In this appendix we briefly define the notations used in the main text for the Lie algebras psl$(2|2)$ and osp$(n|2)$.

\subsection{The Lie superalgebra psl(2$|$2)}

First we consider the notation for psl(2$|$2). The same notation is used for psu(2$|2$) with appropriate realness conditions on the fields.
The Lie superalgebra psl(2$|$2) is generated by six bosonic elements $E^0_i,E^\pm_i$ and
eight fermionic ones $S^\pm_{\alpha,\beta}$, with $i=1,2$ and  $\alpha,\beta = 1,2$.
Here the notation of \cite{GQS} is adopted.
The bosonic subalgebra is sl(2) $\oplus$ sl(2),
\begin{align}
 [ E^0_i , E^\pm_i ] = \pm E^\pm_i ~, \qquad
 [ E^+_i , E^-_i ] = 2 E^0_i ~ .
\end{align}
Among the bosonic and fermionic generators the relations are
\begin{equation}
 \begin{split}
 &[ E^0_1 , S^\pm_{1 \alpha} ] = \pm \tfrac{1}{2} S^\pm_{1 \alpha} ~, \qquad
 [ E^0_1 , S^\pm_{2 \alpha} ] = \pm \tfrac{1}{2} S^\pm_{2 \alpha} ~,  \\
 &[ E^0_2 , S^\pm_{1 \alpha} ] = \pm \tfrac{1}{2} S^\pm_{1 \alpha} ~, \qquad
 [ E^0_2 , S^\pm_{2 \alpha} ] = \mp \tfrac{1}{2} S^\pm_{2 \alpha} ~,  \\
 &[ E^\pm_1 , S^\mp_{1 \alpha} ] = \pm S^\pm_{2 \alpha}\ ~, \qquad
 [ E^\pm_1 , S^\mp_{2 \alpha }] = \mp S^\pm_{1 \alpha}\ ~,  \\
 &[ E^\pm_2 , S^\mp_{1 \alpha} ] = \pm S^\mp_{2 \alpha}\ ~, \qquad
 [ E^\pm_2 , S^\pm_{2 \alpha} ] = \mp S^\pm_{1 \alpha}\ ~.
\end{split}
\end{equation}
The fermionic generators satisfy
\begin{equation}
\begin{split}
  &\{ S^\pm_{1\alpha} , S^\pm_{2\beta} \} = \mp 2 \epsilon_{\alpha\beta}
 E^\pm_1~, \ \qquad \qquad
  \{ S^\pm_{1\alpha} , S^\mp_{2\beta} \} = \pm 2 \epsilon_{\alpha\beta}
 E^\pm_2 ~,  \qquad \ \\
 & \{ S^+_{1\alpha} , S^-_{1\beta} \} = 2 \epsilon_{\alpha\beta}
 (E^0_1 - E^0_2) ~, \qquad
 \{ S^+_{2\alpha} , S^-_{2\beta} \} = 2 \epsilon_{\alpha\beta}
 (E^0_1 + E^0_2) ~.
\end{split}
\end{equation}
The invariant bilinear form is
\begin{equation}
 \begin{split}
 &\text{str} \, E_1^0 E_1^0 = \frac{1}{2}~, \ \ \  \qquad
 \text{str} \, E_1^+ E_1^- = 1 ~, \quad\ \qquad
 \text{str} \, E_2^0 E_2^0 = - \frac{1}{2}~,\ \  \\
 &\text{str} \, E_2^+ E_2^- = - 1 ~, \qquad
 \text{str} \, S^\pm_{1 \alpha} S^{\mp}_{1 \beta}
  = 2 \epsilon_{\alpha \beta} ~, \qquad
 \text{str} \, S^\pm_{2 \alpha} S^{\mp}_{2 \beta}
  = 2 \epsilon_{\alpha \beta} ~,
\end{split}
\end{equation}
with $\epsilon_{12}=1$.

These generators can be expressed by using $4 \times 4$ matrices.
With the usual Pauli matrices
\begin{align}
 &\sigma^3 =
 \begin{pmatrix}
  1 & 0 \\
  0 & - 1
 \end{pmatrix} ~, \qquad
 \sigma^+ =
 \begin{pmatrix}
  0 & 1 \\
  0 & 0
 \end{pmatrix} ~, \qquad
  \sigma^- =
 \begin{pmatrix}
  0 & 0 \\
  1 & 0
 \end{pmatrix}
\end{align}
the bosonic generators are given by
\begin{align}
 E^0_1 = \frac{1}{2}
\begin{pmatrix}
  \sigma^3 & 0 \\
  0 & 0
\end{pmatrix} ~, \qquad
 E^\pm_1 =
\begin{pmatrix}
  \sigma^\pm & 0 \\
  0 & 0
 \end{pmatrix}~, \qquad
 E^0_2 = \frac{1}{2}
\begin{pmatrix}
  0 & 0 \\
  0 &  \sigma^3
 \end{pmatrix} ~, \qquad
 E^\pm_2 =
\begin{pmatrix}
  0 & 0 \\
  0 & \sigma^\pm
 \end{pmatrix}~. \nonumber
\end{align}
Defining the following matrices as
\begin{align}
  \hat \sigma^+ =
 \begin{pmatrix}
  1 & 0 \\
  0 & 0
 \end{pmatrix} ~, \qquad
  \hat \sigma^- =
 \begin{pmatrix}
  0 & 0 \\
  0 & 1
 \end{pmatrix} ~,
\end{align}
fermionic generators%
\footnote{These are generators of SU(2$|$2) Lie algebra, therefore these
generators satisfy the above \mbox{(anti-)}commutation relations for the PSU(2$|$2) 
Lie superalgebra up to the generator of U(1) Lie algebra.}
are given by
\begin{align}
& S^\pm_{11} = \sqrt{2}
\begin{pmatrix}
  0 & \sigma^\pm \\
  0 & 0
\end{pmatrix} ~, \qquad
S^\pm_{12} = \sqrt{2}
\begin{pmatrix}
  0 & 0 \\
  \sigma^\pm & 0
\end{pmatrix} ~, \\
& S^\pm_{21} = \sqrt{2}
\begin{pmatrix}
  0 & \pm \hat \sigma^\pm \\
  0 & 0
\end{pmatrix} ~, \qquad
S^\pm_{22} = \sqrt{2}
\begin{pmatrix}
  0 & 0 \\
  \mp \hat \sigma^\mp & 0
\end{pmatrix} ~. \nonumber
\end{align}

\subsection{The Lie superalgebra osp($n|$2)}

A good way to describe the Lie superalgebra is via supermatrices.
Define the matrix
\begin{align}
 G_{IJ} = \left (
 \begin{array}{c|cc}
  \mathbb{I}_n & 0 & 0 \\ \hline
   0 & 0 & 1 \\
   0 & -1 & 0
 \end{array} \right) ~,
\end{align}
where the label $I,J$ runs from $1$ to $n+2$,
and $\mathbb{I}_n$ is the $n \times n$ identity matrix.
Moreover, we define $(e_{IJ})_{KL} = \delta_{IL} \delta_{JK}$.
Using $i,j = 1, \cdots , n$ and $\bar i , \bar j = n+1, n+2$,
the generators of osp($n|$2) are then given by
\begin{align} \nonumber
 E_{ij} = G_{ik} e_{kj} - G_{jk} e_{ki} ~, \qquad
 E_{\bar i \bar j} = G_{\bar i \bar k} e_{\bar k \bar j}
 + G_{\bar j \bar k} e_{\bar k \bar i} ~ , \qquad
 E_{i \bar j} = E_{\bar j i} = G_{i k} e_{k \bar j}
 + G_{\bar j \bar k} e_{\bar k i} ~.
\end{align}
The sl(2) subgroup is generated by
\begin{equation}
 E_{(n+1)(n+2)} = - 2 E^0 ~, \qquad E_{(n+1)(n+1)} = 2 E^+ ~, \qquad
 E_{(n+2)(n+2)} = - 2 E^-
\end{equation}
and the fermionic part is
\begin{equation}
 E_{i(n+1)} = F^+_i ~, \qquad E_{i(n+2)} = F^-_i ~.
\end{equation}
Commutation relations are
\begin{align}
 &[E^0 ,E^\pm] = \pm E^\pm ~, \qquad [E^+ , E^-] = 2 E^0 ~, \\
 &[E^0 , F_i^\pm] = \pm \tfrac12 F_i^\pm ~ ,\qquad
  \{ F_{i}^\pm , F_{j}^\pm \}
  = \pm 2 \delta_{i,j} E^\pm ~. \nonumber
\end{align}

\section{Extended superconformal algebras}
\label{symmetry}

In this appendix we show that the theories obtained have desired symmetries.
In the next subsection, we study the theory with small ${\cal N}=(4,4)$
superconformal symmetry. In appendix \ref{Kniz} we summarize the Knizhnik's
so$(n)$ superconformal algebra \cite{K}. In appendix \ref{ESA}, we study the theories with general extended superconformal symmetry.

\subsection{Small ${\cal N}=4$ superconformal algebra}

We will now show that the theory with action \eqref{eq:N4action} provides a free field realization of small ${\cal N}=4$ superconformal algebra.
The small ${\cal N}=4$ superconformal algebra is generated
by bosonic currents $T(z),J^i(z)$ with $i=1,2,3$ and
fermionic currents $G_\alpha^\pm$ with $\alpha = 1,2$.
The central charge is $c = 6 (k-1)$ and the level of the SU(2) currents
is $k-1$.
The currents have the following OPEs
(see, for instance, \cite{deBoer:2008ss})
\begin{equation}
\begin{split}
 T(z) T(w)\ &\sim\ \frac{c/2}{(z-w)^4} + \frac{2 T(w)}{(z-w)^2} +
 \frac{\partial T(w)}{z-w}  ~, \\
 J^i(z) J^j (w)\ &\sim\ \frac{k-1}{2}\frac{ \delta^{ij}}{(z-w)^2} +
  \frac{ i \epsilon^{ijk} J^k (w)}{z-w} ~, \\
 T(z) J^i (w)\ &\sim\ \frac{J^i (w)}{(z-w)^2} +  \frac{\partial J^i (w)}{z-w} ~, \\
 T(z) G_\alpha^\pm (w)\ &\sim\ \frac{3}{2}\frac{ G_\alpha^\pm (w)}{(z-w)^2} +  \frac{\partial G_\alpha^\pm  (w)}{z-w} ~, \\
J^i(z) G_\alpha^\pm (w)\ &\sim\ \pm \frac{1}{2} \sigma^{i,\pm}_{\beta \alpha } \frac{ G_\beta^\pm  (w)}{z-w} ~, \\
 G^+_\alpha (z) G^-_\beta (w)\ &\sim\
  \frac{2c/3 \delta_{\alpha \beta }}{(z-w)^3} + \frac{ 4 \bar \sigma^i_{\alpha \beta} J^i (w)}{(z-w)^2}
   + \frac{2 T(w) \delta_{\alpha \beta} + 2 \bar \sigma^i_{\alpha \beta} \partial J^i (w) }{z-w} ~.
\end{split}
\end{equation}

Let us construct these currents using the symmetries of the action of small $\mathcal N=4$ super Liouville theory in \eqref{eq:N4action}. The field content is the bosonic field $\phi(z)$, the su(2) currents $K^i(z)$
and the fermionic fields $\psi_\pm^\alpha(z)$ with $\alpha = 1,2$.
The OPEs of these fields are
\begin{equation}
\begin{split}
 &\phi (z) \phi (w)\ \sim\ - \frac{1}{2} \ln (z - w) ~, \qquad
  \psi_\pm^\alpha (z) \psi_\mp^\beta (w)\ \sim\ \frac{\delta_{\alpha \beta}}{z-w} ~, \\
& K^i (z) K^j (w)\ \sim\ \frac{k-2}{2}\frac{ \delta^{ij}}{(z-w)^2} +
  \frac{ i \epsilon^{ijk} K^k (w)}{z-w} ~.
\end{split}
\end{equation}
We find that the generators of ${\cal N}=4$ super conformal algebra are realized by
\begin{equation}
\begin{split}
 &T (z)\  =\ -  \partial \phi \partial \phi + Q \partial^2 \phi -\frac{1}{2}
 (\psi_+^\alpha \partial \psi_-^\alpha
+ \psi_-^\alpha \partial \psi_+^\alpha )
+ \frac{1}{k} K^i K^i ~, \\
&J^i (z)\  =\ K^i + \frac{1}{2} \psi_+^\alpha \sigma^i_{\alpha \beta} \psi_-^\beta ~, \\
&G^\pm_\alpha\ =\ i \sqrt{2} \left( \psi_\pm^\alpha \partial \phi - Q \partial \psi_\pm^\alpha \pm
  \frac{1}{\sqrt{k}} ( \psi_\pm^\beta \sigma^{i,\pm}_{ \beta \alpha} K^i - \psi_\pm^\alpha \psi_+^\beta \psi_-^\beta )\right)~.
\end{split}
\end{equation}
Here $Q = \sqrt{k} - 1/\sqrt{k}$ and
the normal ordering is implicitly assumed. We have defined $\sigma^{i,+}_{\alpha \beta} = \sigma^{i}_{\alpha \beta}$
and $\sigma^{i,-}_{\alpha \beta} = \bar \sigma^{i}_{\alpha \beta}$ for the simplicity of expressions.

The next step is to show that the term $S_{\text{int}}$ in \eqref{eq:N4action} is a screening charge.
Let us write the supercharges in notation of \cite{Brunner:2006tc}. The relation is
\begin{align}\label{}
    G^\pm=\frac{1}{\sqrt{2}}G_1^\pm ~,\qquad G'^\pm_2=-\frac{1}{\sqrt{2}}G^\mp_2 ~,
\end{align}
where $G^\pm$ and $G'^\pm$ both generate an $\mathcal N=2$ superconformal algebra.
The two interaction terms are
\begin{equation}\label{}
     V_1+V_2=\frac{1}{k}  \psi_-  q
 \bar \psi_- e^{ b \phi }
  +\frac{1}{k} \bar \psi_+ q^{-1} \psi_+ e^{ b  \phi } ~,
\end{equation}
the third term in \eqref{eq:N4action} is a contact term that we remove.
We compute
\begin{equation}
\begin{split}
      G^+(z) V_1(w)\ &\sim\  0 ~,\qquad G^-(z) V_1(w) \ \sim\  -\partial\left(\frac{i k^{-1/2}q_{1\alpha}
 \bar \psi_-^\alpha e^{ b \phi }}{z-w}\right) ~, \\
      G'^+(z) V_1(w)\ &\sim\  \partial\left(\frac{i k^{-1/2}q_{2\alpha}
 \bar \psi_-^\alpha e^{ b \phi }}{z-w}\right) ~,\qquad  G'^-(z) V_1(w)\ \sim\ 0
\end{split}
\end{equation}
and
\begin{equation}
\begin{split}
      G^+(z) V_2(w)\ &\sim\  \partial\left(\frac{i k^{-1/2}\bar \psi_+^\alpha q^{-1}_{\alpha 1} e^{ b  \phi }}{z-w}\right)
~,\qquad G^-(z) V_2(w)\ \sim\ 0 ~, \\
      G'^+(z) V_2(w)\ &\sim\  0 ~ ,\qquad
  G'^-(z) V_2(w) \ \sim\ -\partial\left(\frac{i k^{-1/2}\bar \psi_+^\alpha q^{-1}_{\alpha 2} e^{ b  \phi }}{z-w}\right)\, .
\end{split}
\end{equation}
Similarly one can compute that the OPE of the currents $J^i$ and the energy-momentum tensor $T$ with the fields $V_a$
are regular up to total derivatives. Hence, $V_1$ and $V_2$ are screening charges for our free field representation of the
small $\mathcal N=(4,4)$ superconformal algebra.

But they are even more, namely F-terms of $\mathcal N=(2,2)$ superconformal theories,
\begin{equation}\label{eq:fterm}
\begin{split}
  V_1(w) \ &=\ \frac{1}{2\pi i}\oint_w d z\, \Bigl[A G^+(z)(i k^{-1/2}q_{1\alpha}
 \bar \psi_-^\alpha e^{ b \phi })(w)+\\
&\qquad\qquad\qquad +(A-1)G'^-(z)(i k^{-1/2}q_{2\alpha}
 \bar \psi_-^\alpha e^{ b \phi })(w)\Bigr] ~, \\
    V_2(w) \ &=\ -\frac{1}{2\pi i}\oint_w d z\, \Bigl[B G^-(z)(-i k^{-1/2}\bar \psi_+^\alpha q^{-1}_{\alpha 1} e^{ b  \phi })(w)+\\
&\qquad\qquad\qquad +(B+1)G'^+(z)(-i k^{-1/2}\bar \psi_+^\alpha q^{-1}_{\alpha 2} e^{ b  \phi })(w)\Bigr] ~,
\end{split}
\end{equation}
where $A$ and $B$ are free coefficients.
This shows that in the $\mathcal N=(2,2)$ formalism we can write the terms as chiral and anti-chiral F-terms since we also have
\begin{equation}
\begin{split}
 G^-(z)q_{1\alpha}
 \bar \psi_-^\alpha e^{ b \phi }(w)\ &\sim\  0 ~,\qquad G'^+(z)q_{2\alpha}
 \bar \psi_-^\alpha e^{ b \phi }(w) \ \sim \ 0 ~, \\
 G^+(z)\bar \psi_+^\alpha q^{-1}_{\alpha 1} e^{ b  \phi }(w)\ &\sim\ 0 ~ ,\qquad
G'^-(z)\bar \psi_+^\alpha q^{-1}_{\alpha 2} e^{ b  \phi }(w) \ \sim\ 0 ~ .
\end{split}
\end{equation}

\subsection{Knizhnik's ${\cal N}=(n,n)$ superconformal algebra}
\label{Kniz}

The action \eqref{eq:slft} provides a free field realization for Knizhnik's so($n$)-extended superconformal $W$-algebra \cite{K}.
This is generated by so($n$) currents, Virasoro field and super currents.
The so($n$) currents are
\begin{equation}\label{}
    J^a=K^a+\frac{1}{2}\psi^i t^a_{ij}\psi^j\, .
\end{equation}
Here $K^a$ are the so($n$) currents with level $k-1$ and $t^a_{ij}$ are the generators of so($n$).
The supercurrents are
\begin{equation}\label{}
    G^i = i \sqrt{2} \left( \psi^i \partial \phi - Q \partial \psi^i+
  \frac{1}{2\sqrt{k+n/2-2}} ( \psi^j t^{a}_{ij} K^a )\right).
\end{equation}
Note that there is no triple fermion term.
For $n=1,2,3,4$ these generate the $\mathcal N=n$ superconformal algebra, while for larger $n$ the OPEs looks like a $W$-algebra \cite{K}:
\begin{align}\label{}
    G^i(z)G^j(w)\sim \frac{B\delta^{ij}}{(z-w)^3}+\frac{K t^a_{ij} J^a(w)}{(z-w)^2}+\frac{\tfrac{1}{2}K t^a_{ij} \del J^a(w)+2\delta^{ij}T(w)+\gamma\Pi^{ab}_{ij}:J^aJ^b:(w)}{z-w},
\end{align}
where the last term is the standardly normal ordered product of $J^a$ currents. The constants $B,K,\gamma$ are uniquely determined via the Jacobi identities in terms of $k$ and $n$,
\begin{align}\label{}
    B&= K S\ , & K&=\frac{2S+n-4}{S+n-3}\, , \nonumber\\
    \gamma &=\frac{1}{2}\frac{1}{S+n-3} \ , & S&=2k-1 \ .
\end{align}
The generators of the Lie algebra are denoted $t^a_{ij}$ and are normalized such that $\tr(t^at^b)=-2\delta^{ab}$. Finally, we have defined
\begin{align}\label{}
    \Pi^{ab}_{ij}=t^a_{im}t^b_{mj}+t^b_{im}t^a_{mj}+2\delta^{ab}\delta_{ij}\ .
\end{align}

\subsection{Extended superconformal algebras}
\label{ESA}

These theories were studied in \cite{Ito:1992bi}, and we summarize their results.
Let $t^a$ denote a basis of $\g_0$ and $t^\alpha$ of $\g_1$, i.e. roman indices are bosonic and Greek fermionic.
Then the structure constants are
\begin{equation}
[t^a, t^b] \ = \ {f^{ab}}_c t^c ~ , \qquad
[t^a, t^\alpha] \ = \ {R^{a\alpha}}_\beta t^\beta ~ , \qquad
[t^\alpha, t^\beta] \ = \ {R^{\alpha\beta}}_a t^a ~.
\end{equation}
$R$ indicates that the fermions form a representation of the bosonic subalgebra.

The extended superconformal algebra is generated by
$J^a, T, G^\gamma$ where $\gamma$ is a root in $\Delta_+^1$.
The non-trivial OPEs are
\begin{align}
&J^a(z)G^\gamma(w) \ \sim \ \frac{{R^{a\gamma}}_\beta G^\beta(w)}{(z-w)} ~,
\nonumber \\
&G^{\nu-\gamma'}(z)G^\gamma(w) \ \sim \ \frac{f_1(k)}{k+h^\vee}\frac{\delta_{\gamma',\gamma}}{(z-w)^3}
-\frac{1}{k+h^\vee}\sum_{i=1}^2 f_2^i(k)\Bigl(\frac{2J^i_{\gamma,\gamma'}(w)}{(z-w)^2}+
\frac{\del J^i_{\gamma,\gamma'}(w)}{(z-w)}\Bigr)+ \nonumber \\
&\qquad \qquad \qquad \qquad - \frac{2}{\nu^2}\delta_{\gamma,\gamma'}\frac{ T(w)-S_{\sub}(w)}{(z-w)}+
\frac{1}{k+h^\vee}\frac{(J^2)_{\gamma,\gamma'}(w)}{(z-w)} ~ .
\end{align}
Here $i=1,2$ corresponds to the simple components of $\sub$ and
\begin{align}
J_{\gamma,\gamma'} \ = \ -{R^{\gamma,-\gamma'}}_aJ^a ~, \qquad
(J^2)_{\gamma,\gamma'} \ = \ {R^{\gamma,-\gamma''}}_a \delta_{\gamma''+\gamma''',0}{R^{-\gamma'\gamma'''}}_b    J^a J^b ~.
\end{align}
The $f_i(k)$ are given in table 6 of \cite{Ito:1992bi} and $S_{\sub}$ is a rescaled Sugawara field for the supersymmetric WZNW of the Lie group of $\sub$, see (63) in \cite{Ito:1992bi}.

A free field realization is given by a current algebra of $\sub$, free fermions and the Liouville field $\phi$.
The level of the current algebra is $k_0$ with
\begin{equation}
2(k+h^\vee) = \alpha_L^2 (k_0+h^\vee_0) ~ ,
\end{equation}
 where $h^\vee_0$ is the dual Coxeter number of $\sub$  and $\alpha_L$ is the longest root of  $\sub$.
 The Sugawara energy-momentum tensor is denoted by $\hat T$. 
The free fermions $\psi_\gamma$ transform in the representation $\g_{1/2}$ of $\sub$, i.e. the index $\gamma$ is a root in $\Delta_+^1$.
The Liouville field has energy-momentum tensor
\begin{equation}
T_\phi \ = \ -\del\phi\del\phi+Q\del^2\phi ~ ,
\end{equation}
where
\begin{equation}
Q\ = \ (1-N_f/2)b+b^{-1} ~, \qquad b\ = \ \epsilon\sqrt{\frac{\nu^2}{2(k+h^\vee)}} ~,\qquad N_f\ = \ |\Delta_+^1| ~ .
\end{equation}
Here there is a freedom to choose a sign $\epsilon=\pm$.
The current algebra has standard OPE, for the fermions we choose a basis such that
\begin{equation}
\psi_\gamma(z)\psi_{\nu-\gamma'}(w) \ \sim \ \frac{\delta_{\gamma,\gamma'}}{(z-w)} ~ .
\end{equation}
They form also a current algebra of level being the eigenvalue of the quadratic Casimir in the representation. The currents are
\begin{equation}
j^a \ = \ \frac{1}{2} R^{a\gamma'\gamma}\psi_\gamma\psi_{\nu-\gamma'} ~ .
\end{equation}
The energy momentum tensor is
\begin{equation}
T_\psi \ = \  -\frac{1}{2}\sum_{\gamma\in\Delta^1_+}\psi_{\nu-\gamma}\del\psi_{\gamma} ~ .
\end{equation}
The extended superconformal algebra is then generated by the total energy-momentum tensor, total currents and the supercurrents as
\begin{equation}
\begin{split}
T \ &= \  \hat T_{\sub}+T_\psi+T_\phi ~, \\
J^a \ &= \ \hat J^a + j^a ~, \\
G_\gamma \ &= \ \frac{k+\nu^2/2}{k+h^\vee}\del\psi_\gamma -b(k+h^\vee)\del\phi\psi_\gamma
-\frac{1}{(k+h^\vee)} \sum_{\gamma'}\sum_{i=1}^2\bigl(\hat J^i_{\gamma,\gamma'}+c_2^ij^i_{\gamma,\gamma'}\bigr)\psi_{\gamma'} ~ .
\end{split}
\end{equation}
The number $c^i$ is in table 5 of \cite{Ito:1992bi}.
The anti-holomorphic side is constructed analogously.
The screening charge for this system is \cite{Ito:1992bi}
\begin{equation}\label{eq:screening}
S\ = \ e^{b\phi}\psi_\gamma \Phi^{\gamma,\bar\gamma} \bar\psi_{\bar\gamma} ~,
\end{equation}
where $\Phi$ transforms as $\Delta^1_\pm$ under holomorphic (plus) and anti-holomorphic (minus) currents, and $\bar\gamma$ is in $\Delta^1_-$. This is exactly the form of our fermion interaction term $e^{b\phi}\langle\psi,\Ad(q)\bar\psi\rangle$ in \eqref{eq:generalsuperliouville}.

\end{document}